\documentclass[letterpaper,11pt]{article}
\usepackage{amsmath, amsthm, amssymb}
\usepackage{bridges}
\usepackage{graphicx}
\usepackage[colorlinks=true, urlcolor=blue, citecolor=black, linkcolor=black]{hyperref}
\usepackage{subcaption}
\usepackage{xcolor}
\definecolor{darkgreen}{RGB}{0, 100, 0}
\usepackage{tikz}
\usepackage{float}

\title{Immersive Visualization of Flat Surfaces Using Ray Marching}

\author{Fabian Lander\textsuperscript{1} and Diaaeldin Taha\textsuperscript{2}
\vspace{10pt}\\
Max Planck Institute for Mathematics in the Sciences, Leipzig, Germany\\
\textsuperscript{1}{lander@mis.mpg.de}, \textsuperscript{2}{taha@mis.mpg.de}
}
\date{}

\begin{document}

\maketitle

\thispagestyle{empty}

\begin{abstract}
We present an effective method for visualizing flat surfaces using ray marching. Our approach provides an intuitive way to explore translation surfaces, mirror rooms, unfolded polyhedra, and translation prisms while maintaining computational efficiency. We demonstrate the utility of the method through various examples and provide implementation insights for programmers. Finally, we discuss the use of our visualizations in outreach. We make our simulations and code available online.
\end{abstract}

\begin{figure}[ht]
    \centering
    \includegraphics[width=\textwidth]{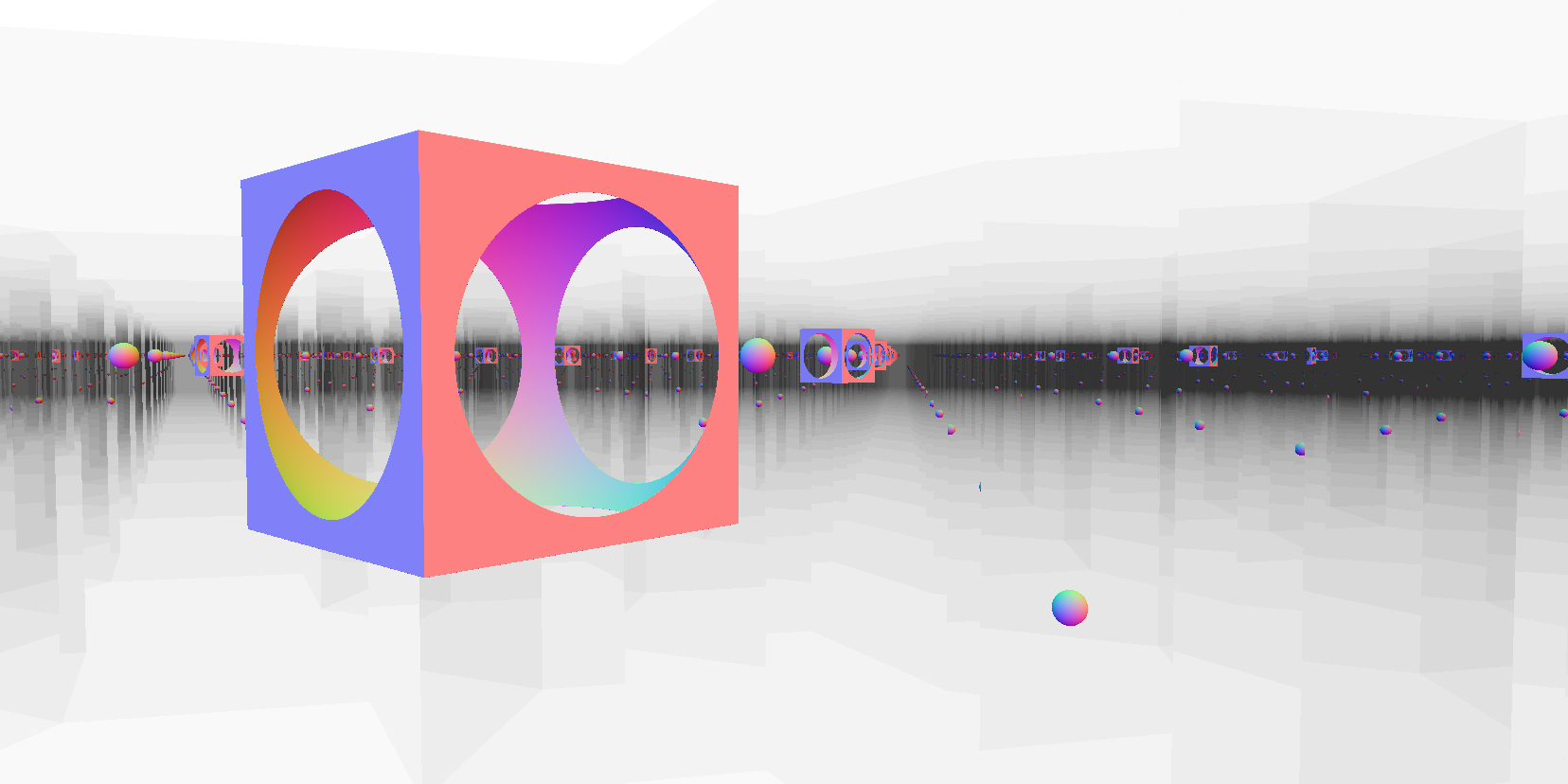}
    \caption{A first-person view inside a ``room'' whose base is the double pentagon flat surface.}
    \label{fig:cover}
\end{figure}

\section*{Introduction}

Flat surfaces naturally arise in various mathematical contexts, from paper folding to dynamical billiards. The simplest example of a flat surface is the torus: a rectangle with opposite edges identified by translation. Walking off the right edge returns one to the left, and stepping off the top returns one to the bottom. This method of constructing surfaces by identifying parallel edges generalizes to more complex polygons, giving rise to a wide variety of flat surfaces. For example, Figure~\ref{fig:combined_translation_surface} shows an L-shaped polygon whose sides are identified to form one such surface. At first glance, these surfaces seem simple, as they are locally isometric to the Euclidean plane, but their global structure and dynamics can be surprisingly complex.

\begin{figure}[ht]
    \centering
    \includegraphics[width=\textwidth]{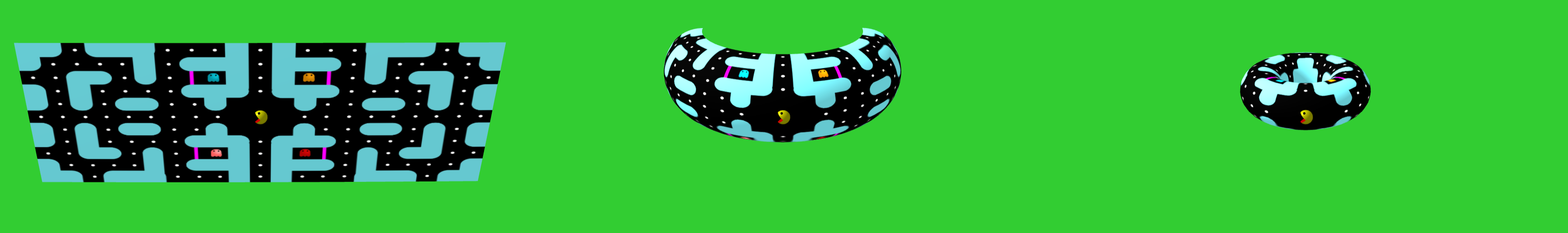}
    \caption{Gluing opposite sides of a rectangle forms the simplest example of a flat surface: the torus. Though the scales are faithful, the torus appears smaller than expected because its radius equals the rectangle's height. For a magnified view from a different angle, see Appendix~\ref{app:magnified_torus}.}
    \label{fig:torus}
\end{figure}

In this work, we visualize flat surfaces by rendering how they appear from the perspective of a viewer who inhabits a thickened three-dimensional version of the surfaces, offering an intuitive first-person experience of their intrinsic geometry. We build on a large body of work on rendering geometric spaces in virtual reality, including the works of Weeks \cite{weeks2002real,weeks2006real}, Nelson-Segerman \cite{nelson2017visualizing}, Hart-Hawksley-Matsumoto-Segerman \cite{bridges2017:33,bridges2017:41}, Coulon-Matsumoto-Segerman-Trettel \cite{bridges2020:153, bridges2020:161,coulon2022ray}, and many others. Additionally, we draw inspiration from Edwin Abbott's \emph{Flatland} and the mirror rooms of Leonardo da Vinci and Yayoi Kusama \cite{applin2012yayoi}.

\section*{Flat Surfaces}

Flat surfaces lie at the intersection of geometry, dynamics, and algebra. Their study has led to deep mathematical breakthroughs connecting complex dynamics, Teichm\"{u}ller theory, ergodic theory, and number theory. Not only are mathematicians captivated by their rich structure, but these surfaces have also captured the public imagination through, e.g., their connection to billiards. Indeed, the flow of a billiard ball on a polygonal table can be unfolded to a geodesic on a flat surface, providing an elegant bridge between recreational mathematics and cutting-edge research. This connection has helped illuminate both the highly regular and chaotic behaviors that can emerge from seemingly simple geometric systems.

In what follows, a flat surface \cite{zorich2006flat} will refer to a surface obtained by taking a finite collection of polygons $\{P_1,\ldots,P_n\}$ in $\mathbb{R}^2$ and identifying their edges pairwise by isometries (translations, rotations, and reflections), resulting in a surface $X$ with a flat metric everywhere except possibly at a finite set of singular points $\Sigma$ (corresponding to some vertices of the polygons after gluing) where the cone angle may differ from $2\pi$. Note that not all vertices necessarily become singularities after gluing, as exemplified by the standard torus construction. The metric on $X \setminus \Sigma$ is induced from the Euclidean metric on $\mathbb{R}^2$.

We note that our visualization method can be readily extended to affine surfaces as well, where instead of using only isometries, we allow edges to be identified by invertible affine transformations.  These surfaces carry a well-defined affine structure, although they generally do not inherit a global flat metric. Our approach can also handle certain simple infinite flat surfaces, such as infinite flat cones and cylinders.

\section*{Ray Marching}

Ray marching is a class of graphics rendering methods where each pixel on the screen emits a ray into the scene, and that ray is traversed in small steps rather than solving for explicit intersection points. This approach dates back to at least the 1980s, including early work by Perlin-Hoffert \cite{perlin1989hypertexture}. Many variants of ray marching exist. For instance, \emph{volume rendering} \cite{drebin1988volumerendering} samples density at each step, while \emph{sphere tracing} \cite{hart1996sphere} relies on signed distance functions (SDFs) to determine a safe step size.

Ray marching is particularly suited for visualizing complex geometric spaces, such as flat surfaces. It only requires local 
distance checks and straightforward step updates, with no need to solve complicated
intersection equations. Additionally, the method maps well to modern GPUs,
enabling real-time interactive visualizations of fractals, non-Euclidean rooms,
and other exotic geometries.

A signed distance function (SDF) is a function that returns the signed shortest distance $d(\vec{p})$ from a point $\vec{p}$ to the surface of an object. By convention, if $d(\vec{p}) > 0$ then $\vec{p}$ is outside the object, if $d(\vec{p}) = 0$ then $\vec{p}$ lies exactly on its surface, and if $d(\vec{p}) < 0$ then $\vec{p}$ is inside the object. A sphere provides a simple example. For a sphere centered at $\vec{c}$ with radius $r$:
\begin{equation*}
    d_{\text{sphere}}(\vec{p}) =
    \|\vec{p} - \vec{c}\| - r.
\end{equation*}
Beyond this elementary example, the computer-graphics community has developed SDFs for an extensive collection of geometric primitives and complex shapes~\cite{hart1996sphere, quilez2008rendering, quilezSDF}. More complex and organic shapes can be assembled through \emph{constructive solid geometry} (CSG) operations \cite{foley1996computer} such as
\emph{union}, \emph{intersection}, and \emph{difference}, and is something of an art form in its own right. For instance,
the SDF for the difference of a cube and a sphere is
\begin{equation*}
    d_{\Delta}(\vec{p}) = \max\bigl(d_{\text{cube}}(\vec{p}),\,-d_{\text{sphere}}(\vec{p})\bigr).
\end{equation*}
By combining SDFs through these operations, we obtain a single SDF that describe the entire scene.

Each pixel is rendered by casting a ray from the camera into the scene and advancing it step-by-step according to the signed distance function of the scene. The ray direction $\vec{v}$ is computed by normalizing the vector from the camera to the location of the pixel in the view plane. The ray is then advanced iteratively according to the signed distance function of the scene.
We start at the camera position $\vec{p}_0$ and repeatedly update
\begin{equation*}
    \vec{p}_{n+1} = \vec{p}_{n} + d(\vec{p}_n)\,\vec{v},
\end{equation*}
where $d(\vec{p}_n)$ is the distance to the nearest surface at position $\vec{p}_n$.
If $d(\vec{p}_n)$ falls below a small scene-scale-dependent threshold (e.g., $\varepsilon = 10^{-4}$ units of length), we consider the ray to have hit an object and assign the corresponding pixel a color based on the object's material or shading model. If the ray travels beyond a predefined maximum distance without hitting any object, it is assumed to have missed all surfaces, and the pixel is assigned a background color.

\begin{figure}[H]
    \centering
    \begin{subfigure}[b]{0.45\textwidth}
        \centering
        \includegraphics[width=\textwidth]{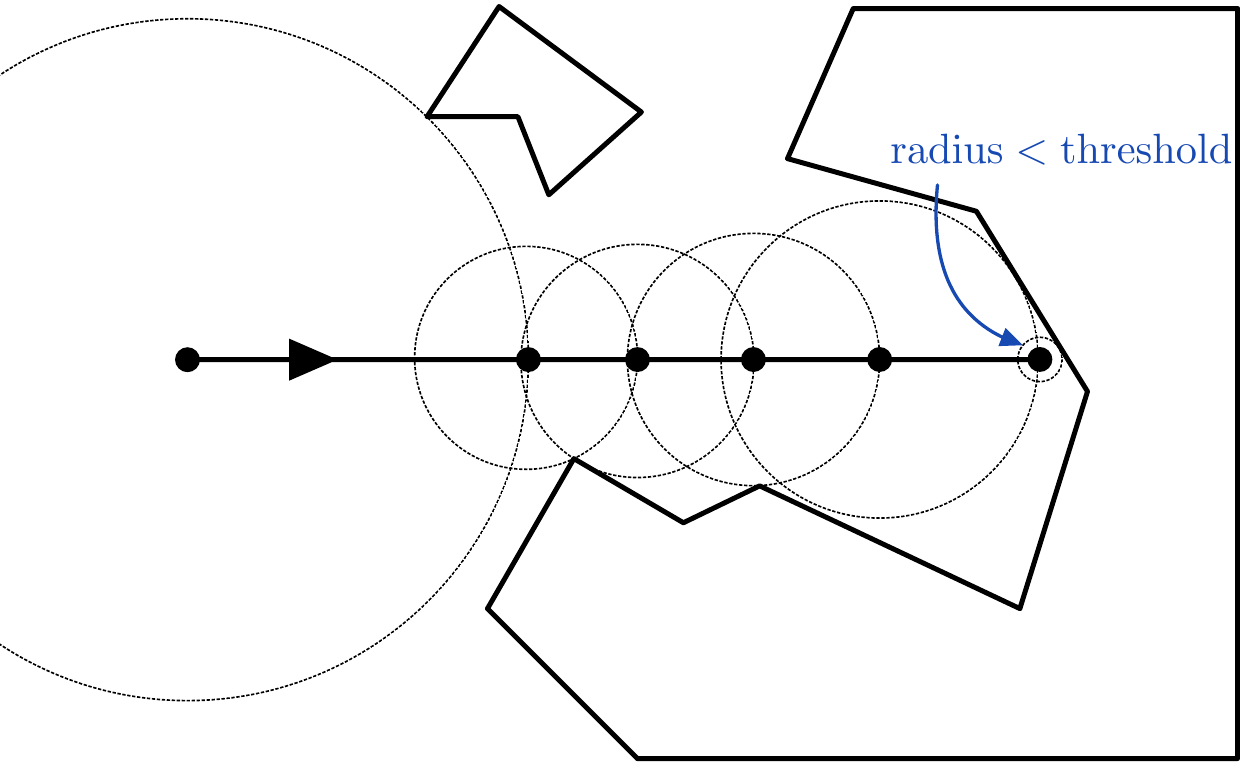}
        \caption{}
        \label{fig:ray_marching}
    \end{subfigure}
    \qquad
    \begin{subfigure}[b]{0.45\textwidth}
        \centering
        \includegraphics[width=\textwidth]{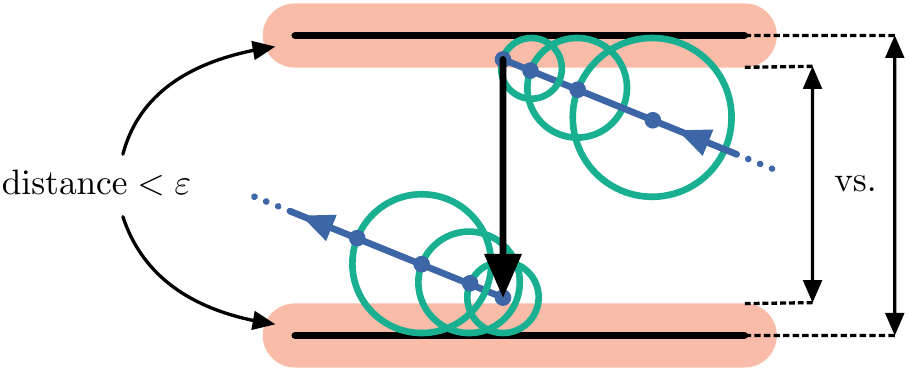}
        \caption{}
        \label{fig:safe_translation}
    \end{subfigure}
    \caption{Illustration of the extended ray marching algorithm. (a) The core ray marching algorithm, where a ray iteratively approaches the surface of an object represented as a signed distance function (SDF). (b) Rays are translated and rotated when they encounter walls, with an offset to prevent immediate reintersection.}
\end{figure}

Once a ray hits an object, the final color at the corresponding pixel can be computed using a desired lighting or shading model. In our simulations, we use a simple coloring based on normal vectors. We approximate $\nabla d(\vec{p})$ at the hit point $\vec{p}$, where $\nabla d(\vec{p})$ is the gradient of the signed distance function of the object and thus points in the direction normal (perpendicular) to the object's surface $d(\vec{p}) = 0$. These normal vectors can then be mapped to RGB color values directly or combined with classic shading models like Blinn–Phong. More advanced methods such as physically based rendering can also be used.

\section*{Ray Marching on Flat Surfaces}

\begin{figure}[h]
    \centering
    \begin{subfigure}[b]{0.3\textwidth}
        \centering
        \includegraphics[width=\textwidth]{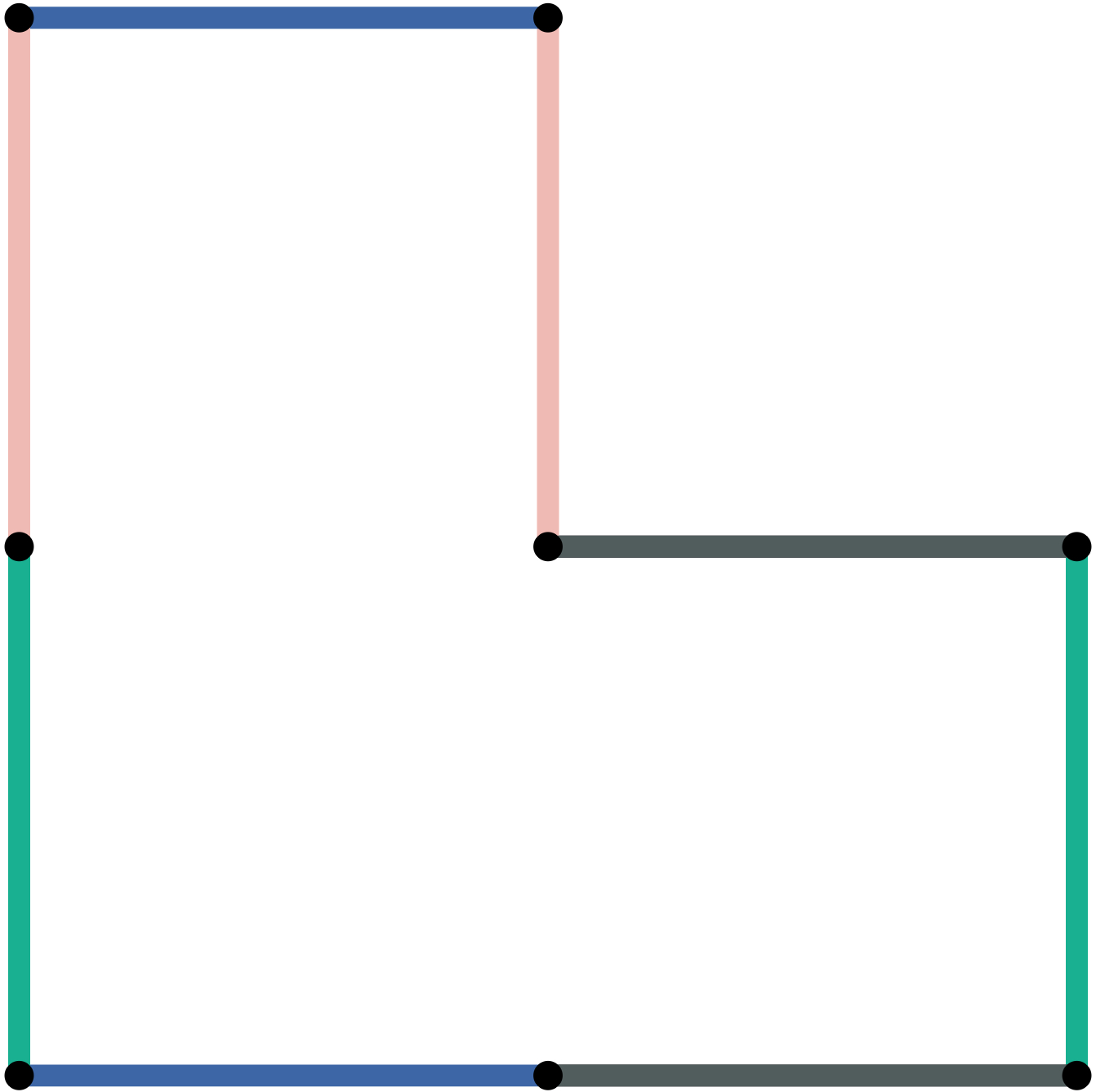}
        \caption{}
        \label{fig:translation_surface_L}
    \end{subfigure}
    \qquad
    \begin{subfigure}[b]{0.3\textwidth}
        \centering
        \includegraphics[width=\textwidth]{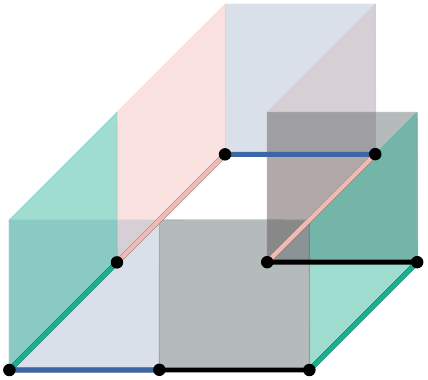}
        \caption{}
        \label{fig:3d_translation_surface}
    \end{subfigure}
    \caption{Visualization of the translation surface $L$ and its three-dimensional thickening. (a) The translation surface $L$ with identified edges marked in matching colors. Parallel opposite edges are identified by translation. (b) Three-dimensional visualization of $L$ with identified opposite walls shown in matching colors.}
    \label{fig:combined_translation_surface}
\end{figure}

We present an adaptation of ray marching to flat surfaces. For our primary example, we consider a specific translation surface $L$, shown in Figure~\ref{fig:translation_surface_L}, whose shape resembles the letter ``L''. The surface is defined by identifying the opposite edges by translations. While we focus on this concrete case, our approach readily generalizes to other flat surfaces such as unfolded polyhedra and mirror rooms.

To create an immersive visualization, we lift $L$ into three dimensions by adding a Euclidean height component to thicken the surface. This transforms our surface into a three-dimensional ``room,'' where parallel walls are identified by translations, leaving the vertical component unchanged. For simplicity, we continue to refer to this three-dimensional space as a ``translation surface'' and use the same notation $L$.

In our ray marching setup, the scene includes both a test object---a cube with a sphere removed---and the walls of $L$, as shown in Figure~\ref{fig:3d_translation_surface}. Rays are traced until they either intersect the test object or encounter a wall. For wall hits, the ray's position is updated based on the gluing rules of the surface.

This process introduces a technical challenge: standard signed distance functions (SDFs) only provide the distance to the nearest surface but do not indicate which object the ray is interacting with. To address this, we augment the SDF with a flag that distinguishes between the walls and the test object.

When a wall is hit, the appropriate translation is applied, and the ray's position is adjusted to avoid immediate re-intersection. Specifically, the translated point is placed at a safe distance greater than $\varepsilon$ from the new wall, as illustrated in Figure~\ref{fig:safe_translation}.

Our implementation targets geometries obtained by thickening a finite collection of Euclidean polygons and assigning boundary rules (translations, rotations, reflections) to the vertical walls and optionally the floors and the ceilings of the thickened polygons:
\begin{itemize}
	\item \emph{Translation surfaces}: matching parallel walls are identified by horizontal translations.
	\item \emph{Unfolded polyhedra}: matching walls are identified by horizontal translations and rotations.
	\item \emph{Mirror rooms}: walls act as ideal mirrors.
	\item \emph{Translation prisms}: translation surface rooms in which the floors and ceilings are identified by a vertical translation, so that the height coordinate becomes periodic.
\end{itemize}
Visualizations and more intuitive descriptions of translation surfaces and unfolded polyhedra are presented in the \emph{Gallery} section, and mirror rooms and translation prisms are presented in the \emph{Extended Gallery} section (Appendix \ref{app:extended-gallery}).

Our interactive visualizations were implemented using WebGL through the three.js framework, which provides an efficient WebGL environment. The visualization interface includes custom JavaScript code for handling keyboard and mouse inputs, allowing users to control the camera and explore the flat surfaces from different perspectives. The core visualization logic, including the ray marching algorithm and the geometric representations of the flat surfaces, was implemented in fragment shaders following the GLSL specification. Our shader‑based implementation can be easily ported to any modern graphics or game engine that supports programmable fragment shaders and user input, such as Unity, Unreal, or Godot. We make our simulations and source code available online~\cite{lander2025raymarchingdemo,lander2025raymarchingrepo}. To run the simulation locally, the code repository can be cloned and hosted as a local server (e.g., using Python’s \texttt{http.server} module).

\section*{Simulation Controls}

Our simulation implements the following controls:
\begin{center}
\begin{tabular}{@{}ll@{}}
\emph{W / A / S / D} & move forward / left / backward / right \\
\emph{E / Q} & move up / down \\
\emph{Mouse}    & rotate the camera
\end{tabular}
\end{center}
We note that the key and mouse mappings mirror standard first-person video game controls, and in our experience, feel immediately intuitive to many users.

\section*{Use in Outreach and Public Engagement}

We use our simulations at science fairs, math festivals, and other public events to illustrate flat surfaces and their surprisingly rich mathematics. We begin with the simplest example: forming a torus by gluing opposite edges of a rectangle, and making a comparison to Pac-Man, where crossing one side of the screen makes one reappear on the other. Next, we highlight more complex surfaces and mirror rooms, drawing parallels to Leonardo da Vinci's mirrored chambers and Yayoi Kusama's infinity mirror rooms. These immersive experiences naturally lead to discussions of ongoing research on flat surfaces and billiards, opening deeper conversations about geometry, topology, and dynamical systems.

\section*{Gallery}

In this work, we present a range of spaces rendered using our ray marching approach, including (thickened) translation surfaces, mirror rooms, (thickened) unfolded polyhedra, and translation prisms. Each example demonstrates how side identifications produce surprisingly immersive and repeated visual effects. We give the walls a subtle glass-like translucency so that viewers can recognize the sides of the base polygons while still seeing the repeating geometry beyond. This translucency creates a faint fog effect in the distance, visible in the rendered images.

\begin{figure}[H]
    \centering
    \begin{subfigure}[b]{0.45\textwidth}
        \centering
        \includegraphics[width=\textwidth]{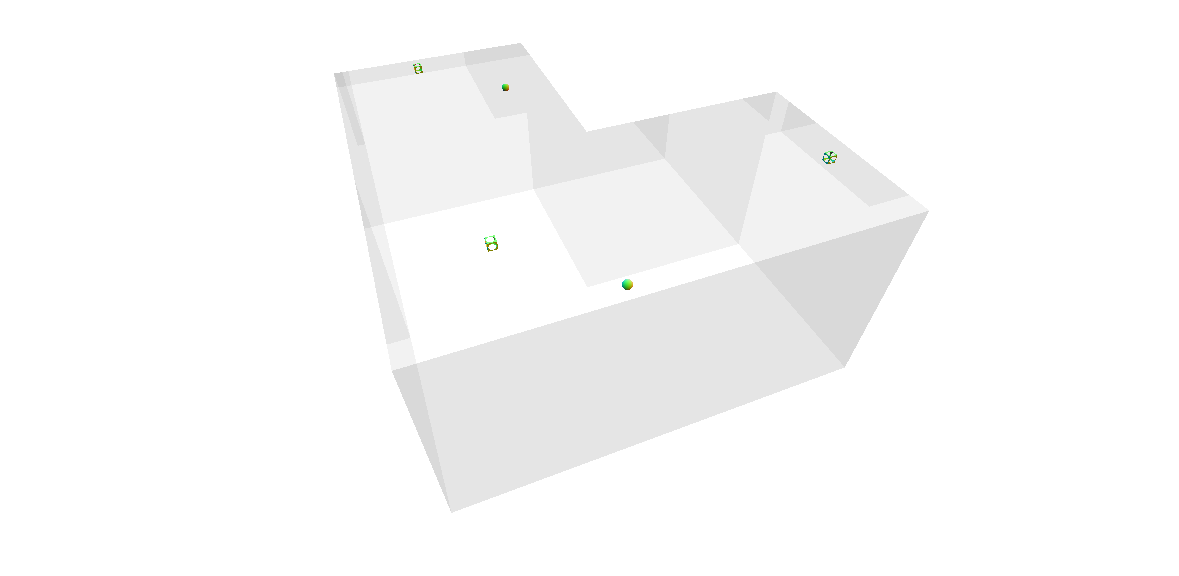}
        \caption{}
        \label{fig:trans1}
    \end{subfigure}
    \hfill
    \begin{subfigure}[b]{0.45\textwidth}
        \centering
        \includegraphics[width=\textwidth]{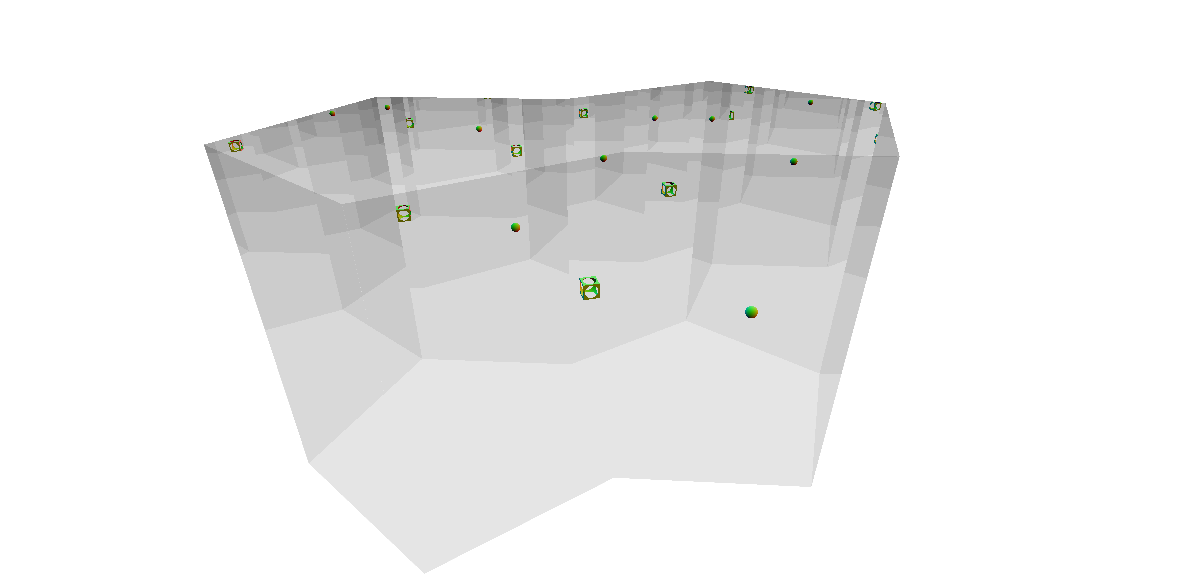}
        \caption{}
        \label{fig:trans3}
    \end{subfigure}
    
    \vspace{1em}
    
    \begin{subfigure}[b]{0.45\textwidth}
        \centering
        \includegraphics[width=\textwidth]{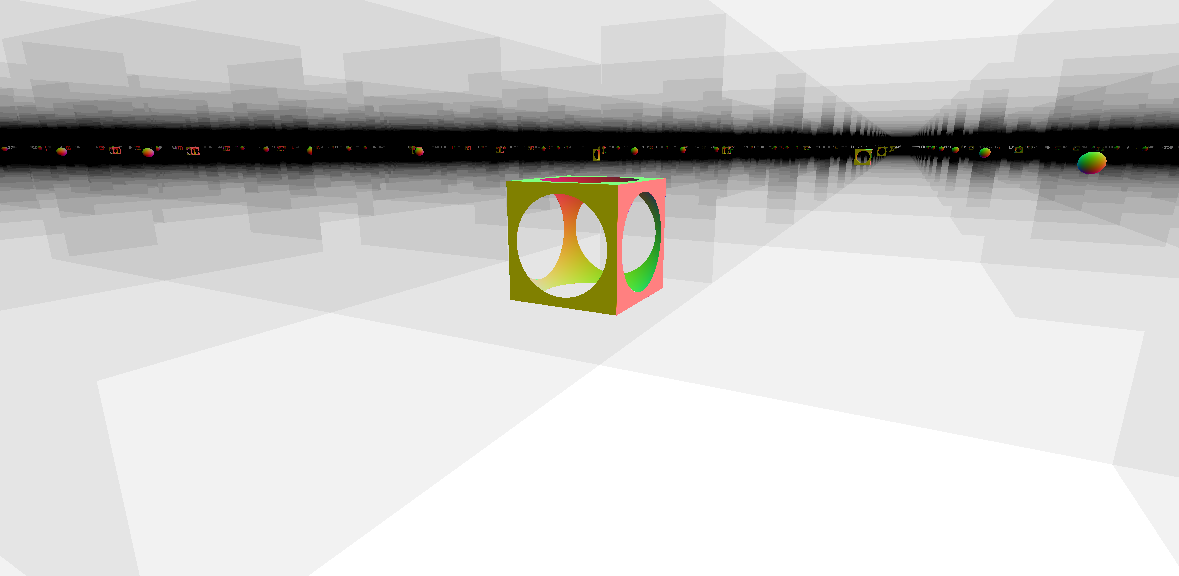}
        \caption{}
        \label{fig:trans2}
    \end{subfigure}
    \hfill
    \begin{subfigure}[b]{0.45\textwidth}
        \centering
        \includegraphics[width=\textwidth]{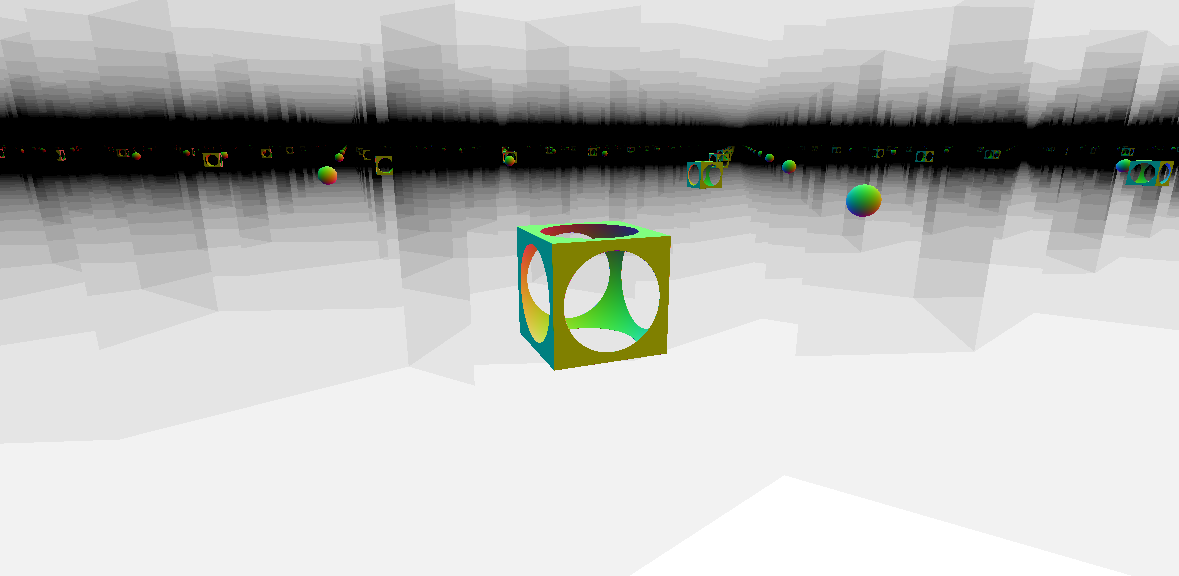}
        \caption{}
        \label{fig:trans4}
    \end{subfigure}
    
    \vspace{1em}
    
    \begin{subfigure}[b]{0.45\textwidth}
        \centering
        \includegraphics[width=0.5\textwidth]{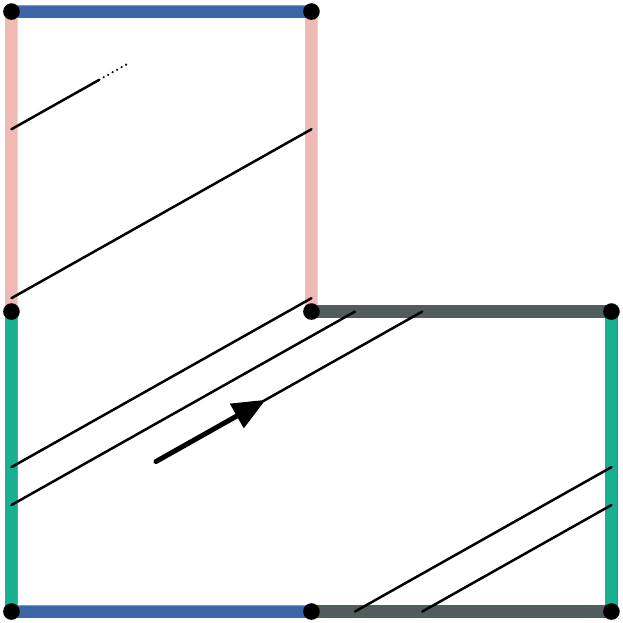}
        \caption{}
        \label{fig:l_shape_floor_plan}
    \end{subfigure}
    \hfill
    \begin{subfigure}[b]{0.45\textwidth}
        \centering
        \includegraphics[width=0.8\textwidth]{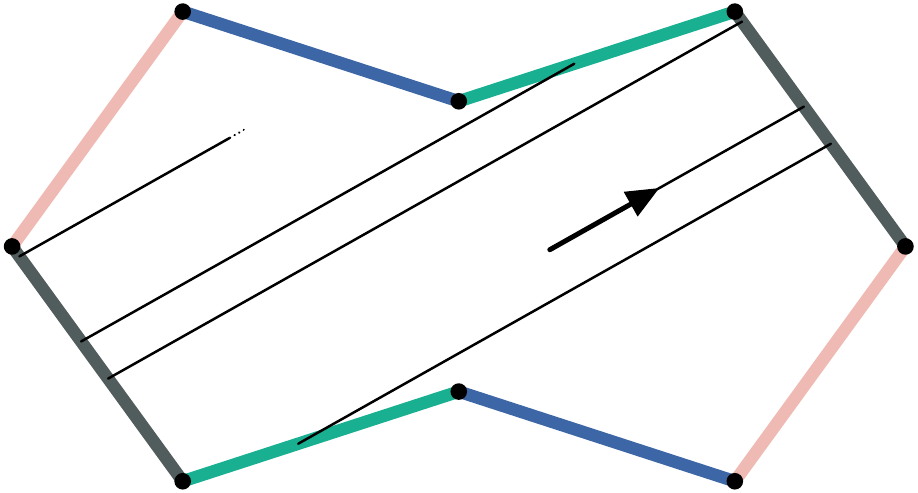}
        \caption{}
        \label{fig:double_pentagon_floor_plan}
    \end{subfigure}
    
    \caption{Visualization of translation surfaces. (a) Top-down view of a simple L-shape. (b) Top-down view of a double pentagon. (c) Inside view of a simple L-shape. (d) Inside view of a double pentagon. (e) Floor plan of the L-shape showing the path of a light ray. (f) Floor plan of the double pentagon showing the path of a light ray.}
    \label{fig:translation_gallery}
\end{figure}

\begin{figure}[H]
    \centering
    \begin{subfigure}[b]{0.45\textwidth}
        \centering
        \includegraphics[width=\textwidth]{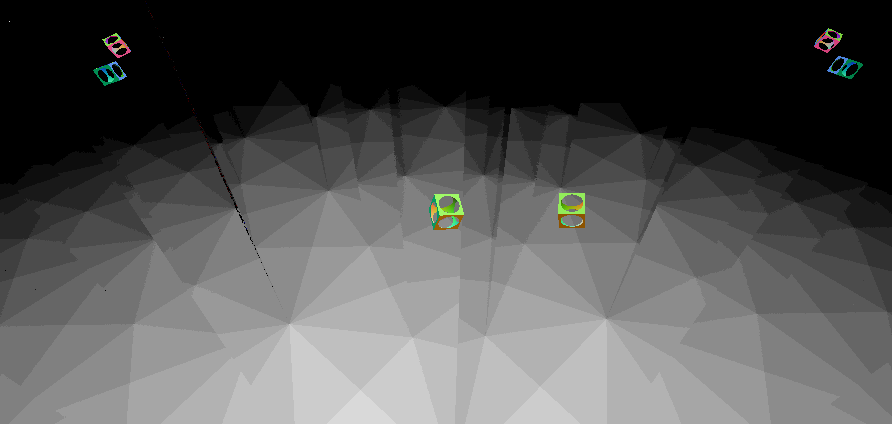}
        \caption{}
        \label{fig:mirror1}
    \end{subfigure}
    \hfill
    \begin{subfigure}[b]{0.45\textwidth}
        \centering
        \includegraphics[width=\textwidth]{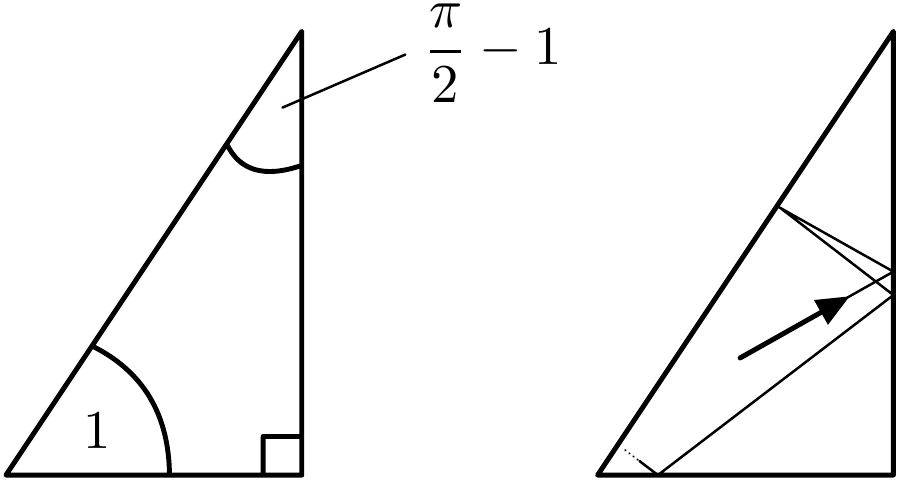}
        \caption{}
        \label{fig:mirror_triangle_floor}
    \end{subfigure}
    \caption{Visualization of a mirror room with a nearly irrational triangle base. (a) First-person view inside the mirror room. (b) Floor plan showing the base triangle and part of a path of a light ray.}
    \label{fig:mirror_gallery}
\end{figure}

\begin{figure}[H]
    \centering
    \begin{subfigure}[b]{0.45\textwidth}
        \centering
        \includegraphics[width=\textwidth]{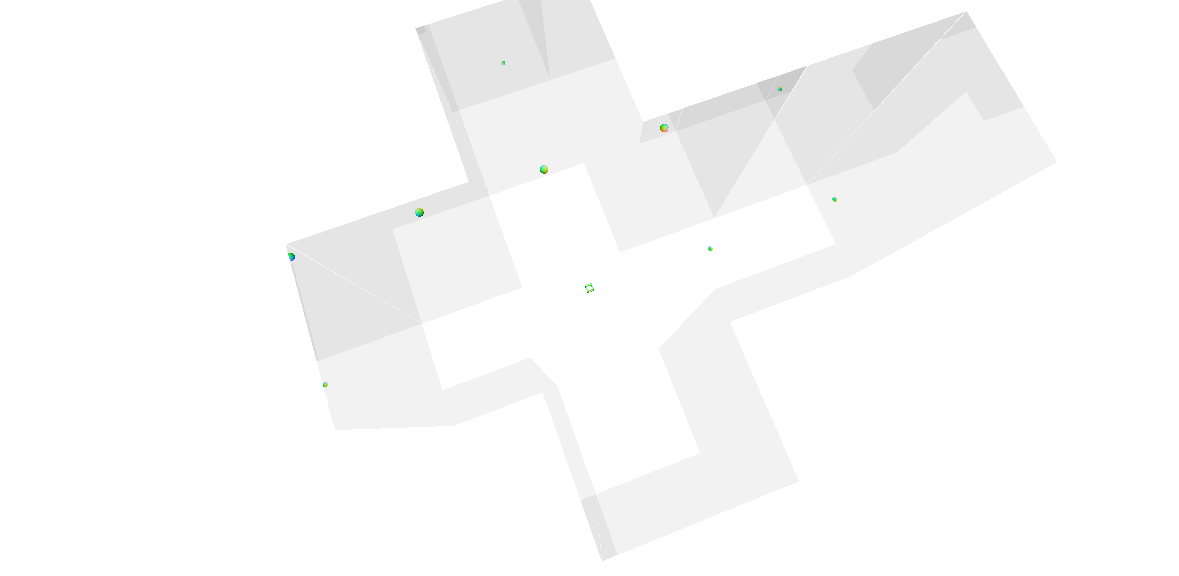}
        \caption{}
        \label{fig:poly1}
    \end{subfigure}
    \hfill
    \begin{subfigure}[b]{0.45\textwidth}
        \centering
        \includegraphics[width=\textwidth]{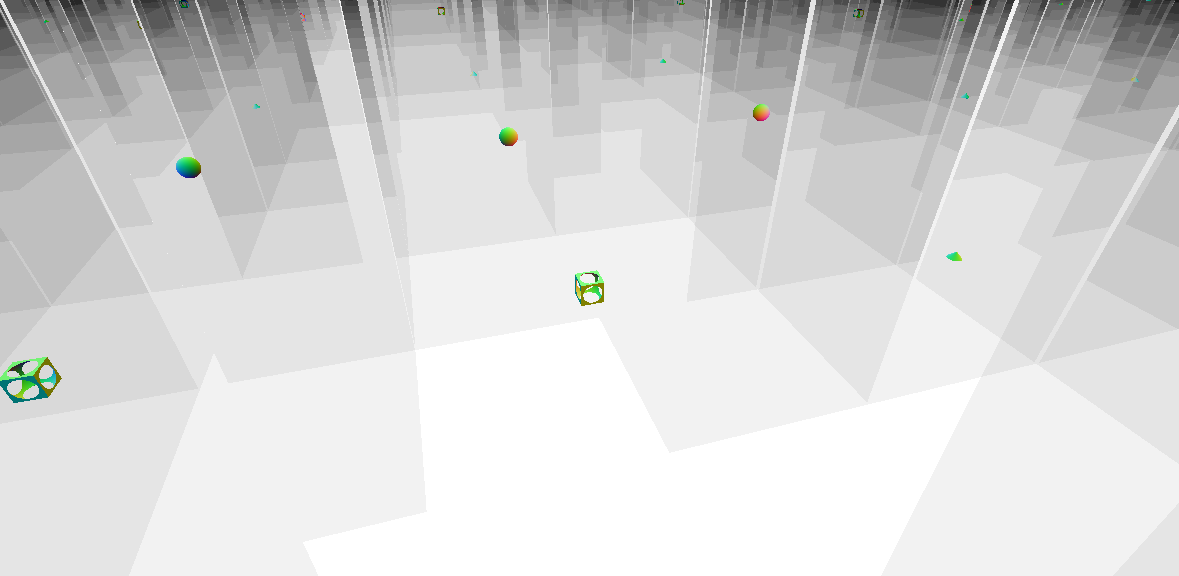}
        \caption{}
        \label{fig:poly2}
    \end{subfigure}
    
    \vspace{1em}
    
    \begin{subfigure}[b]{0.45\textwidth}
        \centering
        \includegraphics[width=0.7\textwidth]{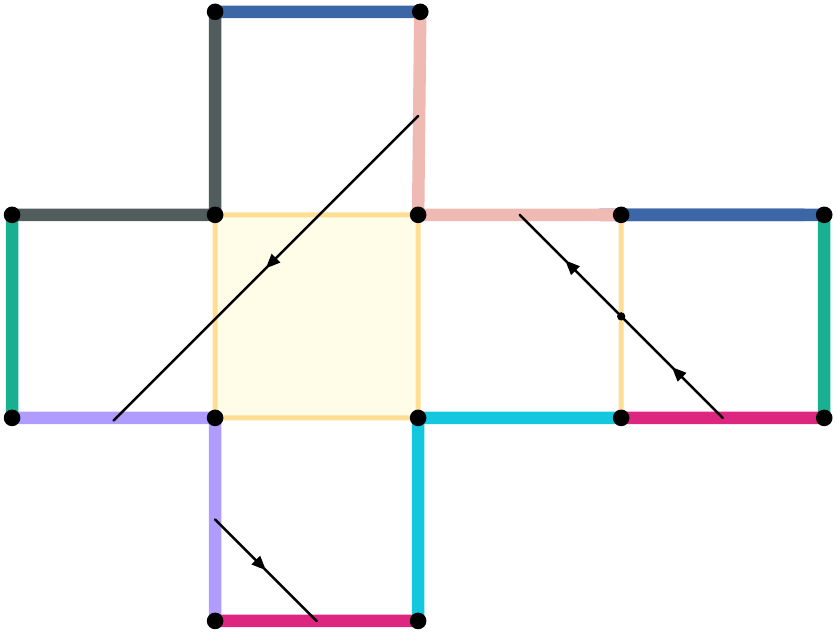}
        \caption{}
        \label{fig:cube_base}
    \end{subfigure}
    \hfill
    \begin{subfigure}[b]{0.45\textwidth}
        \centering
        \includegraphics[width=0.5\textwidth]{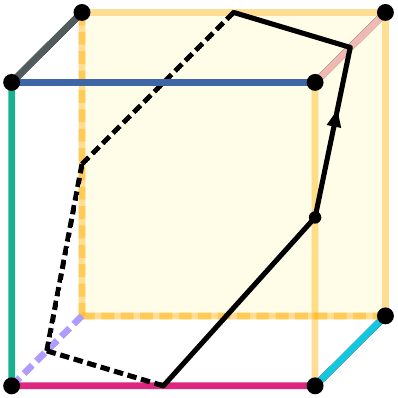}
        \caption{}
        \label{fig:cube_folded}
    \end{subfigure}
    
    \caption{Visualization of an unfolded cube with an added Euclidean height component. (a) Top-down view of the net. (b) First-person perspective inside the unfolded net. (c) Floor plan of the cube net showing how edges are identified and the path of a light ray. (d) The same light ray path shown on the folded cube.}
    \label{fig:polyhedra_gallery}
\end{figure}

We illustrate several representative examples. Figure~\ref{fig:translation_gallery} shows two translation surfaces: an L-shape and a double pentagon. The surfaces are constructed by identifying pairs of parallel sides through translations. Figure~\ref{fig:mirror_gallery} shows a mirror room formed by reflecting rays across the walls of a triangle. Figure~\ref{fig:polyhedra_gallery} shows an unfolded polyhedron arising from flattening the six square faces of a cube and re-gluing matching edges via translations and $90^\circ$ rotations, making it the flat surface analogue of the sphere. Lastly, Appendix~\ref{app:extended-gallery} extends this gallery with more visualizations of \emph{mirror rooms}, \emph{translation prisms}, and \emph{thickened singularities}.

\section*{Summary and Conclusions}

In this work, we use ray marching to visualize flat surfaces, providing an immersive, first-person experience of their intrinsic geometry. Future work could explore several directions: Artistically, real-time interactive installations within these surfaces could bring them to life as engaging public exhibits. Technically, implementing analytic solutions could enhance performance and rendering efficiency. Lastly, extending this work to affine or dilation surfaces would broaden the accessibility of mathematical spaces to the public through immersive visualization.

\section*{Acknowledgements}

This work was carried out at the Heidelberg Experimental Geometry Lab at the University of Heidelberg and the Mathematics Lab at the Max Planck Institute for Mathematics in the Sciences. We thank and acknowledge Mara-Eliana Popescu for her contributions to an earlier version of this work, and R\'{e}mi Coulon for valuable discussions at the beginning of the project.
Figure~\ref{fig:torus} was created by Ricardo Waibel in collaboration with the second author. Fabian Lander was partially supported by the Collaborative Research Center SFB/TRR 191 -
281071066 (Symplectic Structures in Geometry, Algebra and Dynamics) while working on this project.
We sincerely thank the reviewers for their valuable comments and suggestions, which significantly improved the clarity and quality of this manuscript.

{\setlength{\baselineskip}{13pt}
\raggedright
\bibliographystyle{bridges}
\bibliography{references}
}

\clearpage
\appendix

\section{Extended Gallery}
\label{app:extended-gallery}

\begin{figure}[ht]
    \centering
    \begin{subfigure}[b]{0.45\textwidth}
        \centering
        \includegraphics[width=\textwidth]{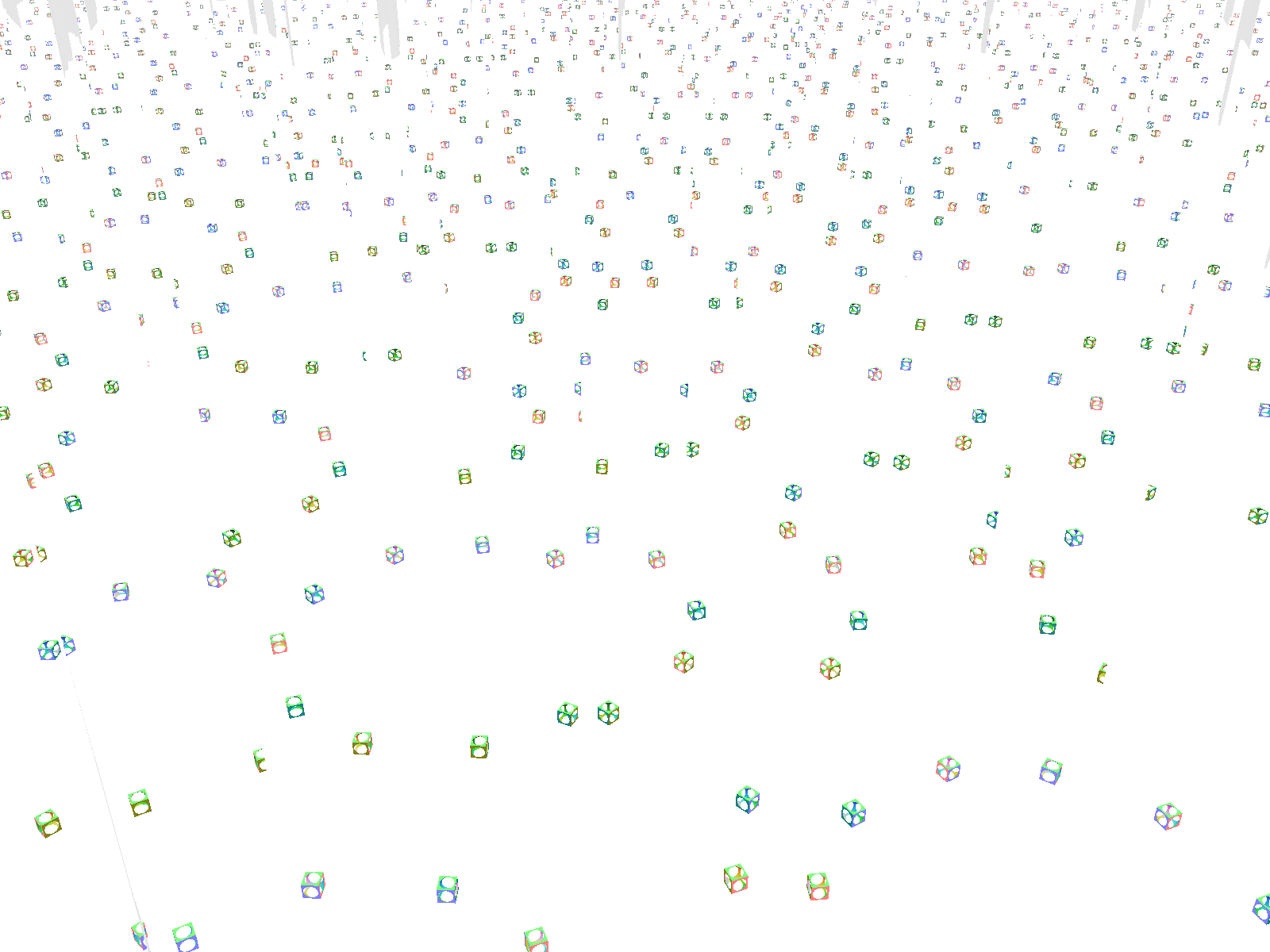}
        \caption{}
        \label{fig:mirror_room1}
    \end{subfigure}
    \hfill
    \begin{subfigure}[b]{0.45\textwidth}
        \centering
        \includegraphics[width=\textwidth]{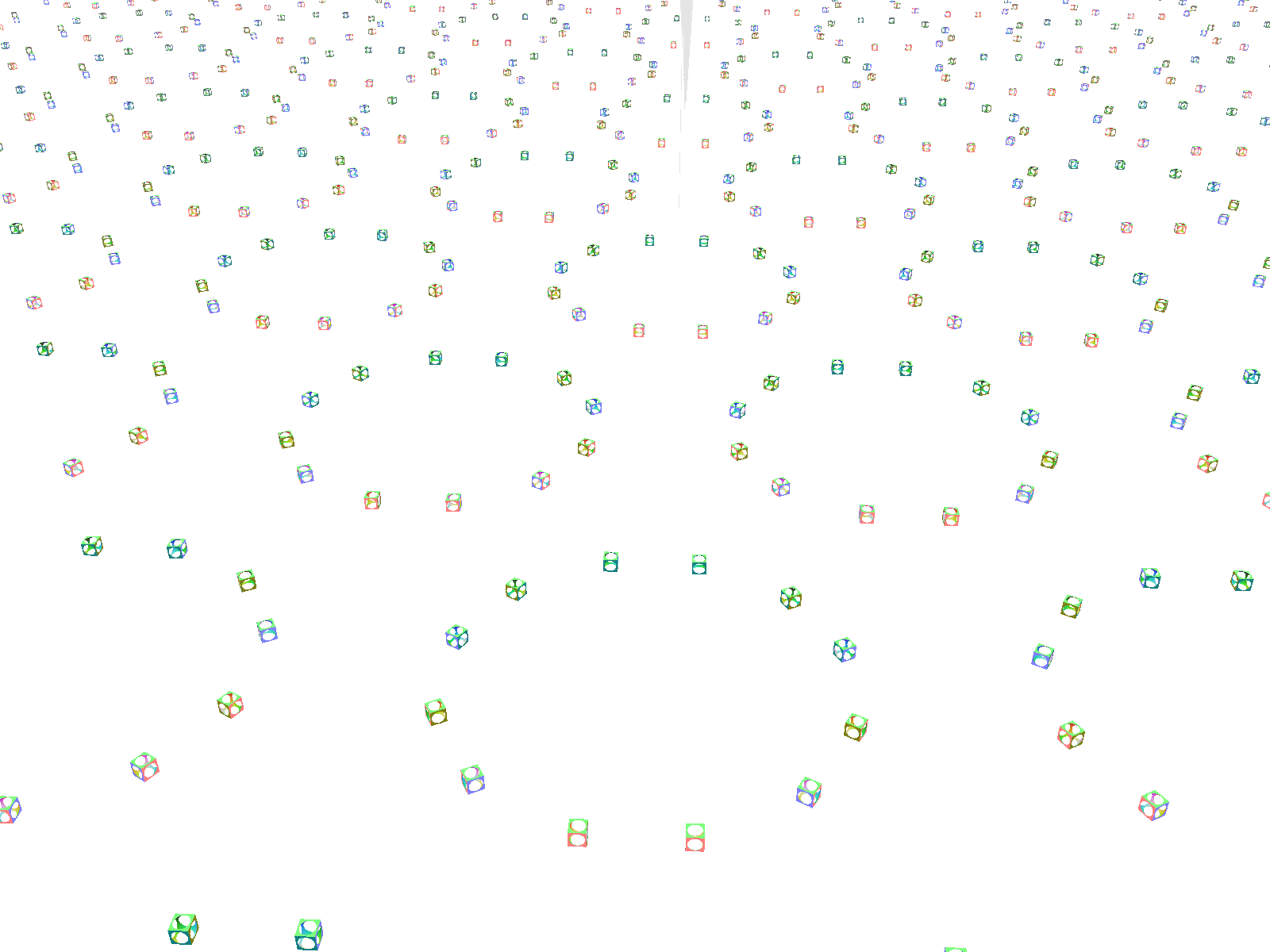}
        \caption{}
        \label{fig:mirror_plan1}
    \end{subfigure}
    
    \vspace{1em}
    
    \begin{subfigure}[b]{0.45\textwidth}
        \centering
        \includegraphics[width=\textwidth]{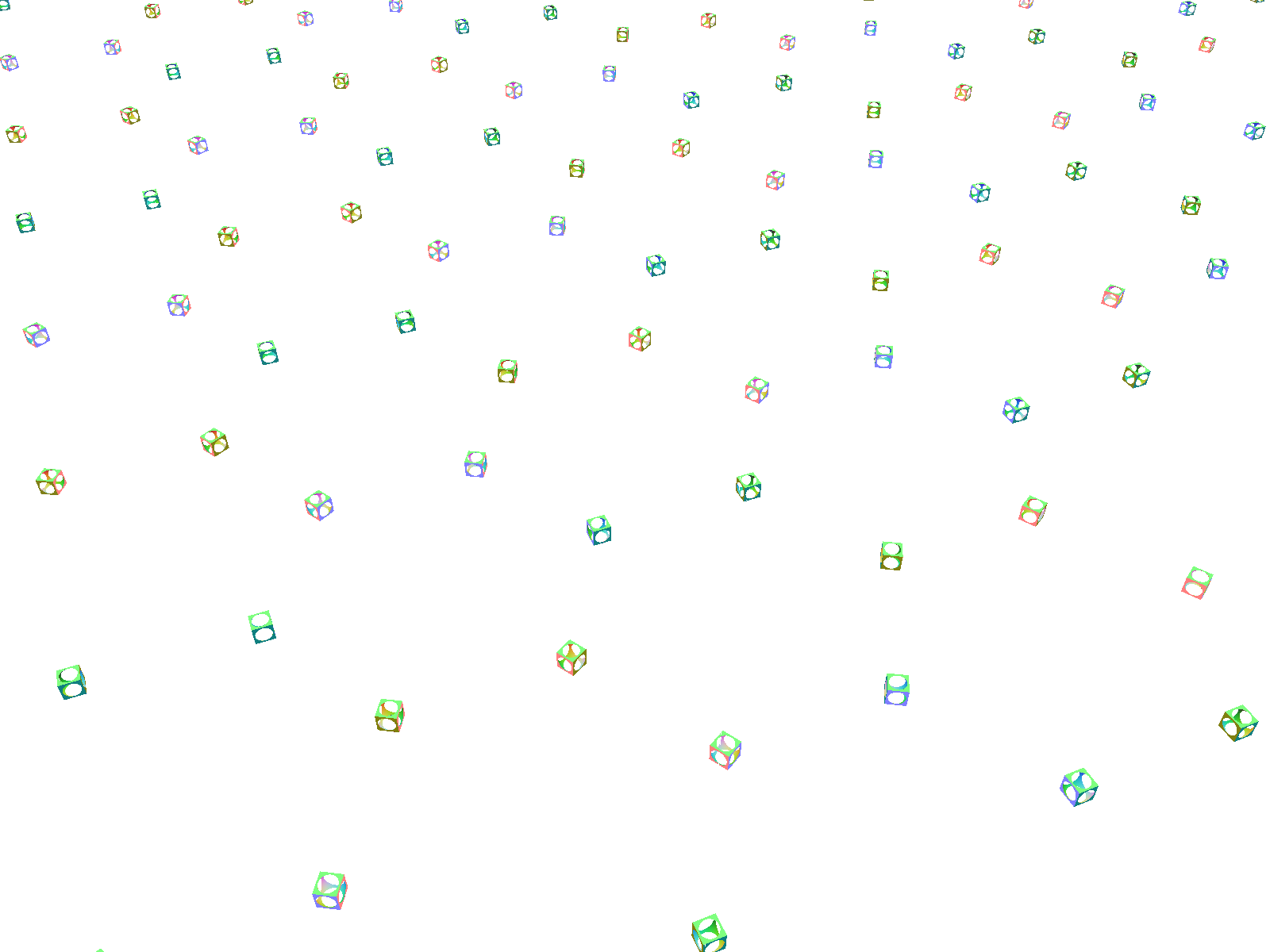}
        \caption{}
        \label{fig:mirror_room2}
    \end{subfigure}

    \caption{Top-down view in mirror rooms with very high walls and whose floor plans are triangles. (a) A triangle with a nearly irrational angle. (b) A $30^\circ$-$60^\circ$-$90^\circ$ triangle. (c) An equilateral triangle.}
    \label{fig:mirror_rooms_gallery}
\end{figure}

\begin{figure}[ht]
    \centering
    \begin{subfigure}[b]{0.45\textwidth}
        \centering
        \includegraphics[width=\textwidth]{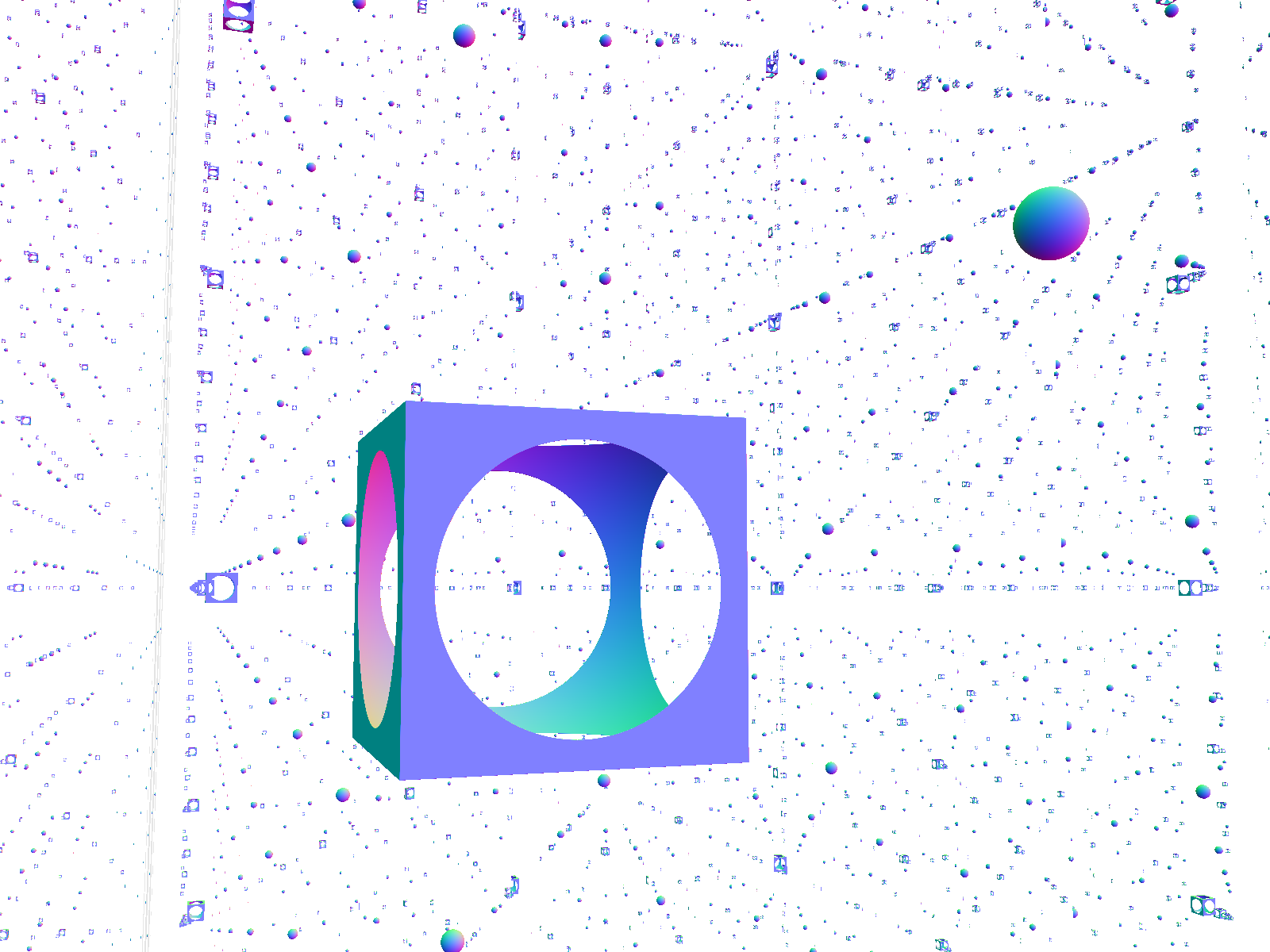}
        \caption{}
        \label{fig:prism1}
    \end{subfigure}
    \hfill
    \begin{subfigure}[b]{0.45\textwidth}
        \centering
        \includegraphics[width=\textwidth]{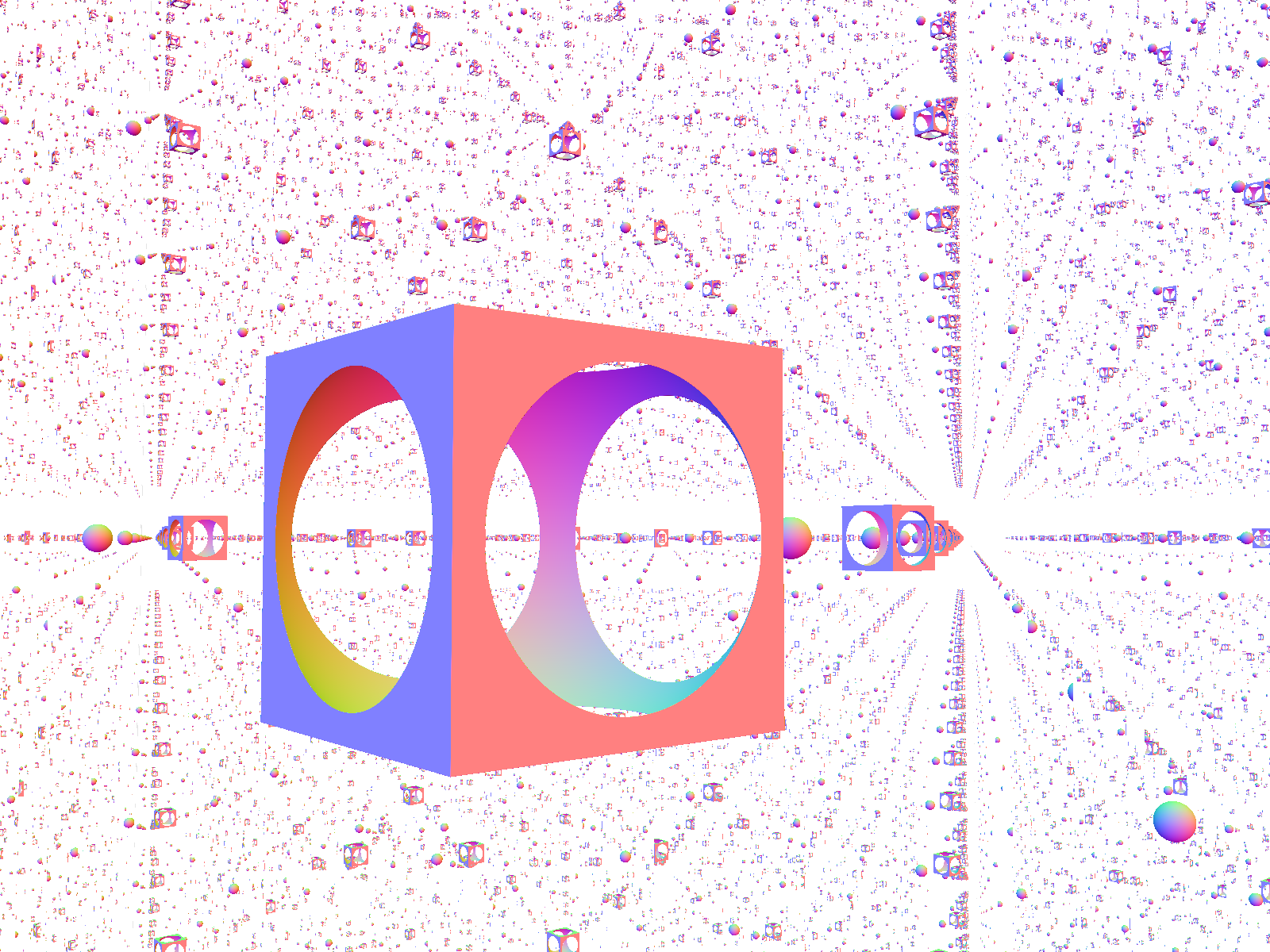}
        \caption{}
        \label{fig:prism2}
    \end{subfigure}
    \caption{Visualizations of translation prisms over the L translation surface and the double pentagon.}
    \label{fig:translation_prism_gallery}
\end{figure}

\begin{figure}[ht]
    \centering
    \begin{subfigure}[b]{0.45\textwidth}
        \centering
        \includegraphics[width=\textwidth]{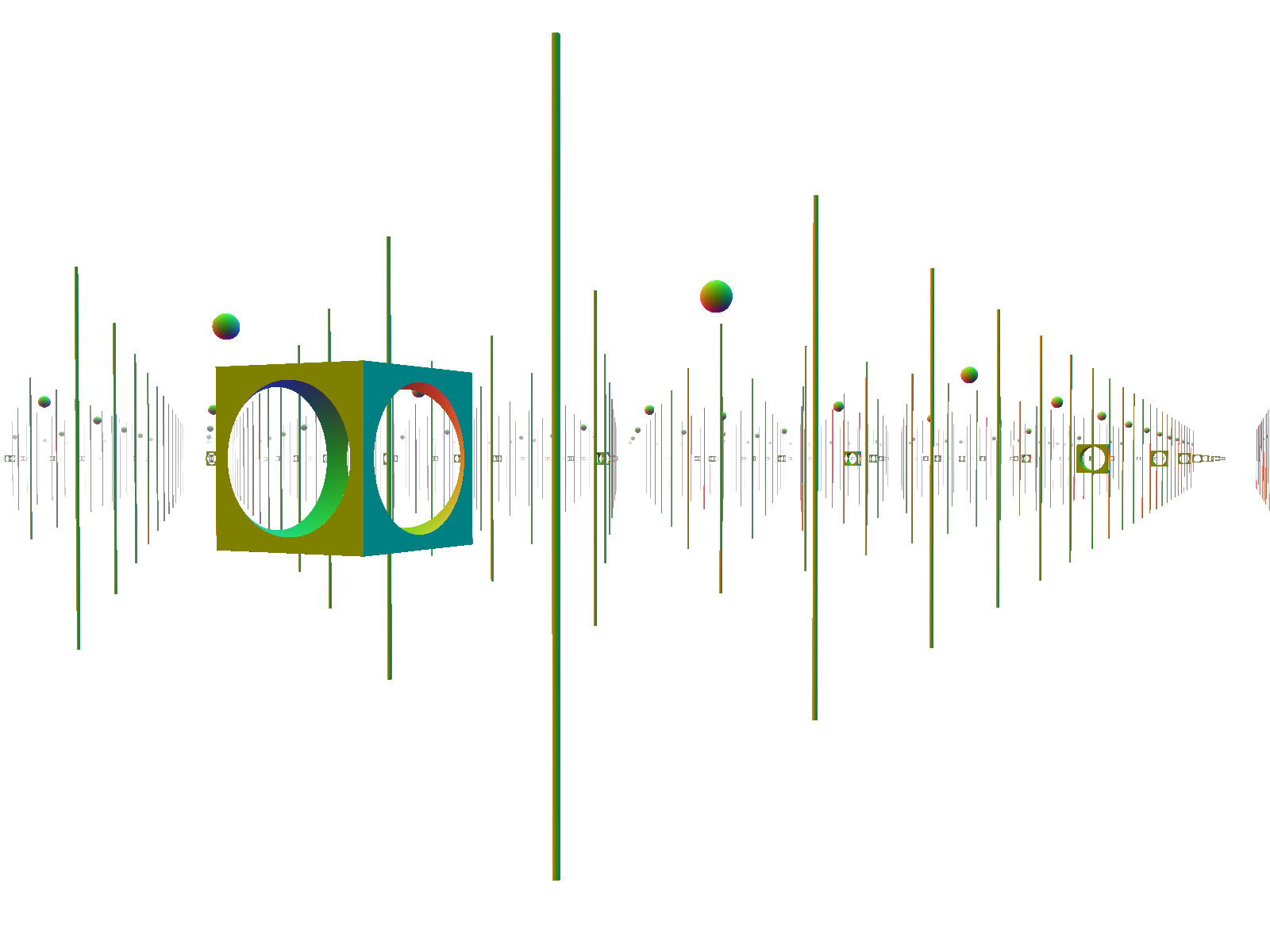}
        \caption{}
        \label{fig:cylinder1}
    \end{subfigure}
    \hfill
    \begin{subfigure}[b]{0.45\textwidth}
        \centering
        \includegraphics[width=\textwidth]{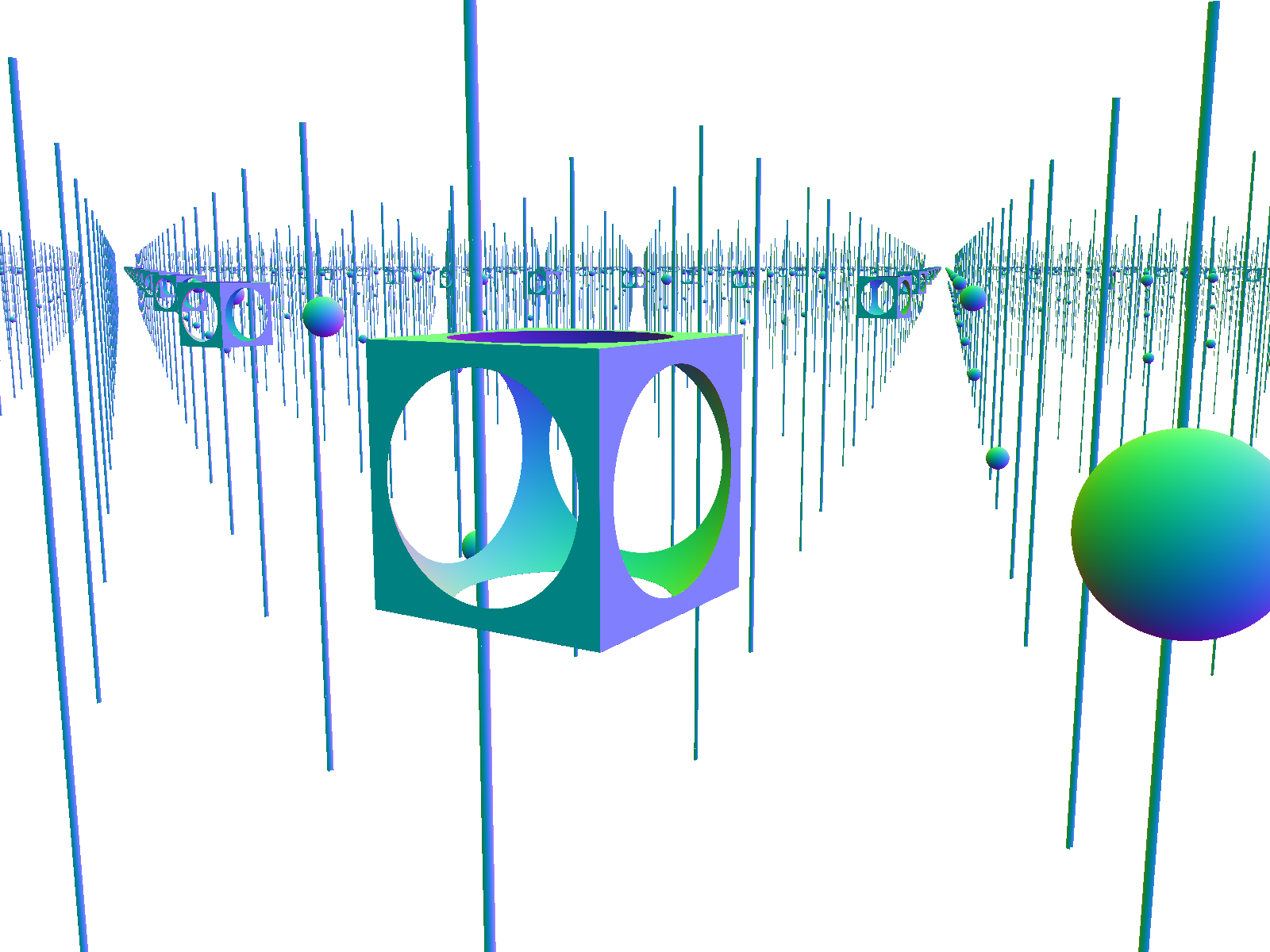}
        \caption{}
        \label{fig:cylinder2}
    \end{subfigure}
    \caption{Visualizations of singular points represented as cylinders in our flat surface renderings.}
    \label{fig:singularity_gallery}
\end{figure}

This appendix presents additional visualizations and intuitive descriptions that complement the examples in the \emph{Gallery} section. We include mirror rooms, translation prisms, and thickened singularity visualizations. Figure~\ref{fig:mirror_rooms_gallery} shows mirror rooms formed by reflecting rays across the walls of triangles. Figure~\ref{fig:translation_prism_gallery} shows two translation prisms \cite{athreya2025linear}, which are closed $3$-manifolds obtained by taking our flat surface rooms and identifying their floors with their ceilings through a vertical translation, so the height coordinate becomes periodic. Figure~\ref{fig:singularity_gallery} shows how singularities in our flat surface visualizations can be rendered as vertical cylinders.

\section{More on Signed-Distance Functions}

A signed distance function (SDF) is a scalar field $d(\vec{p})$ that implicitly encodes a shape by assigning to each point $\vec{p}$ its shortest signed distance to the surface of the object so that the shape itself is recovered exactly as the zero‐level set $d(\vec{p})=0$. SDFs for many primitive shapes, such as spheres, cubes, cylinders, and tori, are well-known and serve as modular building blocks.
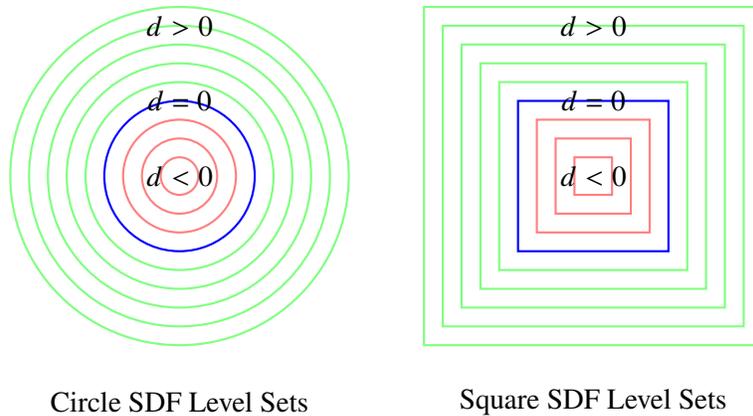
\begin{figure}[ht]
\centering
\begin{tikzpicture}[scale=1.0]
  \begin{scope}
    \foreach \r in {0.25,0.5,0.75} {
      \draw[red!50!white, thick] (0,0) circle (\r);
    }
    \draw[blue, thick] (0,0) circle (1);
    \foreach \r in {1.25,1.5,1.75,2.0,2.25} {
      \draw[green!50!white, thick] (0,0) circle (\r);
    }
    \node at (0,-3) {Circle SDF Level Sets};
    \node at (0,0) {$d < 0$};
    \node at (0,1) {$d = 0$};
    \node at (0,2) {$d > 0$};
  \end{scope}

  \begin{scope}[xshift=5.5cm]
    \foreach \s in {-0.75,-0.5,-0.25} {
      \draw[red!50!white, thick] (-1-\s,-1-\s) rectangle (1+\s,1+\s);
    }
    \draw[blue, thick] (-1,-1) rectangle (1,1);
    \foreach \s in {0.25,0.5,0.75,1.0,1.25} {
      \draw[green!50!white, thick] (-1-\s,-1-\s) rectangle (1+\s,1+\s);
    }
    \node at (0,-3) {Square SDF Level Sets};
    \node at (0,0) {$d < 0$};
    \node at (0,1) {$d = 0$};
    \node at (0,2) {$d > 0$};
  \end{scope}
\end{tikzpicture}
\caption{Two-dimensional signed distance function level sets. Each contour shows points at fixed distance $d$ from the surface: red for $d<0$, blue for $d=0$, and green for $d>0$.}
\label{fig:sdf_2d_levelsets}
\end{figure}
More complex and organic shapes can be constructed by combining and deforming these building blocks using different numerical operations involving the component SDFs, and so that resulting zero-level set matches the intended composite shape. For example, the fundamental set-theoretic operations of union, intersection, and difference translate directly into pointwise $\min$, $\max$, and negation on the SDFs, forming the basis of constructive solid geometry (CSG):
\begin{center}
\begin{tabular}{@{}lll@{}}
\emph{Operation} & \emph{Formula} & \emph{Effect}\\
Union              & $d_\cup(\vec{p}) = \min(d_1(\vec{p}),d_2(\vec{p}))$                       & join two objects\\
Intersection       & $d_\cap(\vec{p}) = \max(d_1(\vec{p}),d_2(\vec{p}))$                       & keep overlap only\\
Difference         & $d_\Delta(\vec{p}) = \max(d_1(\vec{p}),-d_2(\vec{p}))$                     & subtract second object from first\\
\end{tabular}
\end{center}
If a scene consists of $N$ shapes modeled by SDFs $d_1, d_2, \cdots, d_N$, the scene itself is then modeled as one SDF that results from combining all the shapes in the scene:
\begin{equation*}
  d_{\mathrm{scene}}(\vec{p}) = \min_{i=1}^{N} d_i(\vec{p}).
\end{equation*}
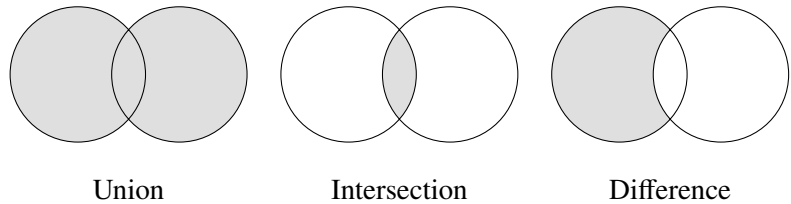
\begin{figure}[h]
  \centering
  \begin{tikzpicture}[scale=0.9]
  \def\r{1}
  \def\dx{1.5}
  \begin{scope}
    \coordinate (C1) at (0,0);
    \coordinate (C2) at (\dx,0);
    \fill[gray!25] (C1) circle (\r) (C2) circle (\r);
    \draw (C1) circle (\r) (C2) circle (\r);
    \node[below] at (0.75, -\r-0.4) {Union};
  \end{scope}
  \begin{scope}[xshift=4.0cm]
    \coordinate (C1) at (0,0);
    \coordinate (C2) at (\dx,0);
    \begin{scope}
      \clip (C1) circle (\r);
      \fill[gray!25] (C2) circle (\r);
    \end{scope}
    \draw (C1) circle (\r) (C2) circle (\r);
    \node[below] at (0.75, -\r-0.4) {Intersection};
  \end{scope}
  \begin{scope}[xshift=8.0cm]
    \coordinate (C1) at (0,0);
    \coordinate (C2) at (\dx,0);
    \fill[gray!25] (C1) circle (\r);
    \begin{scope}
      \clip (C2) circle (\r);
      \fill[white] (C1) circle (\r);
    \end{scope}
    \draw (C1) circle (\r) (C2) circle (\r);
    \node[below] at (0.75, -\r-0.4) {Difference};
  \end{scope}
\end{tikzpicture}
  \caption{The three fundamental CSG operations: \emph{Union} combines objects, \emph{intersection} preserves the overlapping regions, and \emph{difference} subtracts one object from another.}
\end{figure}

For more intricate geometries, practitioners rely on techniques such as voxel-based distance field generation, mesh-to‑SDF conversion algorithms, or, more recently, neural implicit networks trained to approximate continuous SDFs directly from data. We refer the interested reader to the excellent survey \cite{wang20243d}. Lastly, for a comprehensive catalog of hand-crafted primitives, deformation operators, and ready-to-use GLSL code implementations, we refer the reader to Inigo Qu\'ilez’s distance function compendium~\cite{quilezSDF}.

\section{Qualitative Experience and Educational Value}

In our own experience, and in observing users during outreach, we have collected a few consistent observations. First, many abstract properties of flat surfaces come to life in ways that contradict everyday Euclidean expectations: walking straight ahead and, without visual breaks, finding oneself in an identical copy of the room brings the Pac-Man analogy to life, while circling a singularity and finding the room unexpectedly rotated (because the cone angle is not $2\pi$) is surprising the first time it is experienced. We also notice that, after only a few minutes, users start spotting periodic orbits by sight, as periodic orbits pop out as low-level patterns, whereas aperiodic ones appear more visually noisy. Lastly, our simulated mirror rooms are as intriguing as the real ones: when a ray bounces off a side, it extends into another copy of the room that is one reflection away, so reflected objects march off toward vanishing points and shrink at the correct rate, giving an immediate sense of optical ``infinity.'' In short, the simulation lets newcomers absorb new concepts by experiencing the phenomena firsthand, not by reading definitions or staring at static diagrams.

\section{Magnified and Rotated Torus Figure}
\label{app:magnified_torus}

\begin{figure}[ht]
    \centering
    \includegraphics[height=0.85\textheight,keepaspectratio]{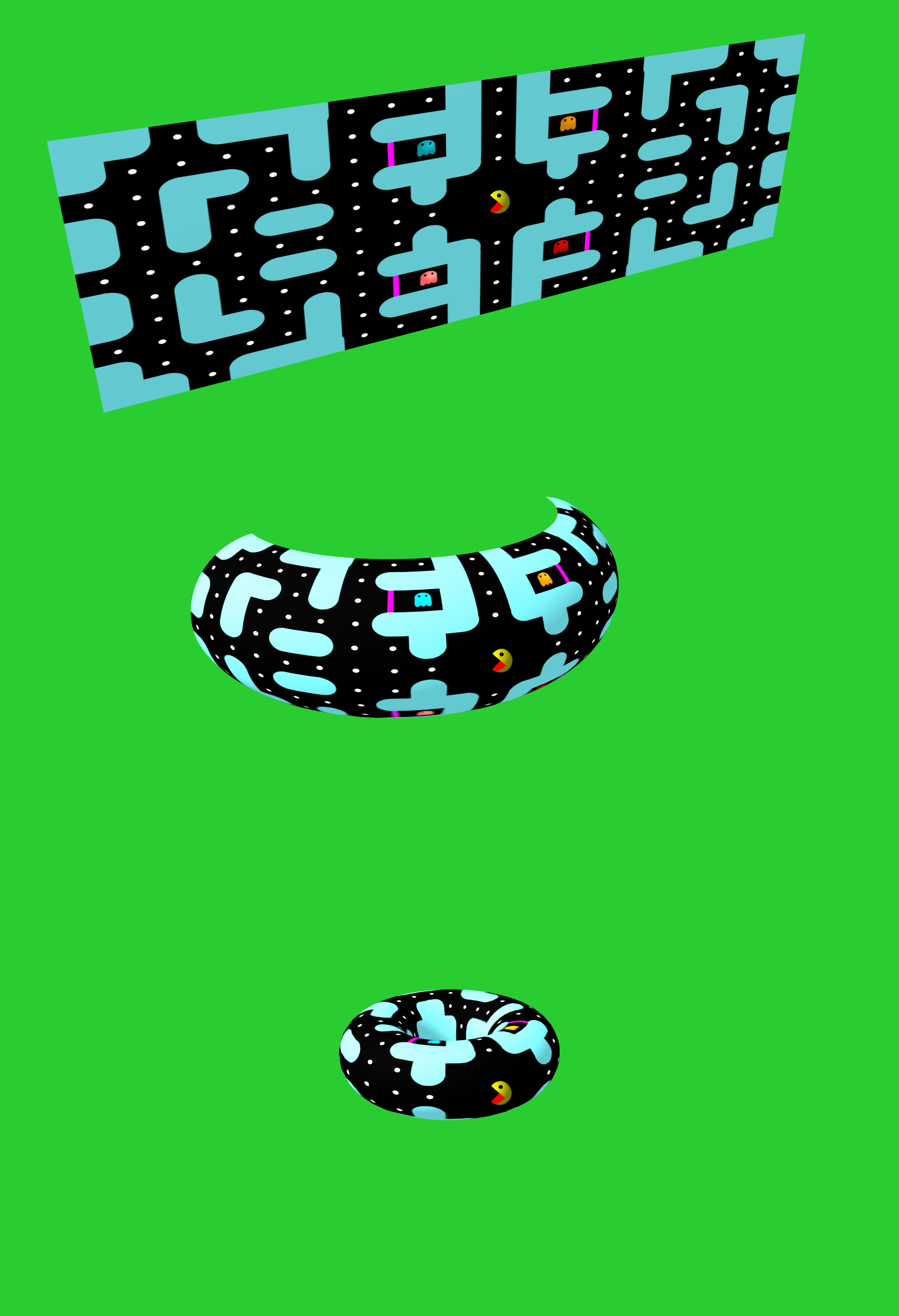}
    \caption{A magnified version of Figure~\ref{fig:torus}.}
    \label{fig:torus_vertical}
\end{figure}

Figure~\ref{fig:torus_vertical} is a magnified view of the torus in Figure~\ref{fig:torus} from a different angle. The geometry is untouched, and the torus is only enlarged and rotated. The apparently ``undersized'' diameter of the torus might appear uncanny at first, yet that scale is the faithful result of rolling the original rectangle into a torus.

\section{Software Design Considerations}

Our software design choices aimed to balance ease of use, the potential for broad adoption, and flexibility by leveraging well-established metaphors and tools:

\begin{itemize}
    \item \emph{Usability}: For end-users, interaction with the simulation is modeled after familiar 3D video game environments. Navigation uses standard first-person and free-camera controls, drawing on well-established, time-tested schemes that users already know or can quickly pick up. Also, the simulation is delivered as a web application which runs directly in modern browsers without requiring any additional installations, plugins, or specific operating systems.
    \item \emph{Adoption}: The implementation uses standard and widely adopted technologies in the computer graphics and mathematical visualization communities:
        \begin{itemize}
        \item Ray marching for rendering signed distance functions, a common technique for visualizing implicit geometries and non-Euclidean spaces.
        \item OpenGL shaders for efficient GPU-accelerated rendering.
        \item Three.js, a popular WebGL-based framework that is accessible, well-documented, and supported by a large community.
        \end{itemize}
    \item \emph{Generality}: Our implementation consists of a compact ray marching kernel and a boundary-identification layer that supports flat spaces defined by edge identifications. Instead of requiring developers to hand-code SDFs directly, users can specify walls, gluing rules, and scene assets using a JSON file, and our system then synthesizes a shader that incorporates these specifications. Each rendered scene in our app is described by a JSON file, making it easy to modify or add examples that fit our framework.
\end{itemize}

\end{document}